\documentclass[pra,aps,twocolumn,amsmath,amssymb,superscriptaddress]{revtex4-1}
\usepackage{epsfig,amsmath}
\usepackage{subfigure}
\usepackage{graphicx}
\usepackage{dcolumn}
\usepackage{stmaryrd}
\usepackage{mathrsfs}
\usepackage{pifont}
\usepackage{amsthm}
\usepackage{amssymb}
\usepackage{bm}
\usepackage{latexsym}
\usepackage[colorlinks=true,linkcolor=cyan,citecolor=cyan]{hyperref}
\usepackage{color}
\usepackage{epstopdf}
\usepackage[ruled]{algorithm2e}

\newcommand{\ket}[1]{\left\vert #1 \right\rangle}
\newcommand{\bra}[1]{\left\langle #1 \right\vert}

\begin{document}

\title{Generating Fock-state superpositions from coherent states by selective measurement}

\author{Chen-yi Zhang}
\affiliation{School of Physics, Zhejiang University, Hangzhou, Zhejiang 310027, China}

\author{Jun Jing}
\email{Email address: jingjun@zju.edu.cn}
\affiliation{School of Physics, Zhejiang University, Hangzhou, Zhejiang 310027, China}

\date{\today}

\begin{abstract}
Fock states and their superpositions are exotic testbeds for nonclassical physics and valuable resources for quantum technologies. We provide a simple protocol for the quantum measurement to generate an arbitrary Fock state and certain superposed Fock states from a coherent state of a target resonator, without any carefully tailored driving. This conditional protocol can be efficiently constructed by a sequence of joint free evolution of the resonator and an ancillary qubit, which are coupled via a Jaynes-Cummings interaction, and projective measurements on the qubit. By properly choosing the duration of each evolution-measurement cycle and the initial state of the resonator, we can generate a desired Fock state $|n\rangle$ and a superposed Fock state $(|0\rangle+|n\rangle)/\sqrt{2}$, $n\sim10$, with a fidelity over $99\%$ in less than $30$ measurements. Moreover, our protocol can be extended straightforwardly to the generation of a Bell-like state $(|00\rangle+|nn\rangle)/\sqrt{2}$ with multiple excitations in a double-resonator system. We also calculate the outcome fidelity and the success probability of our protocol in the presence of decoherence.
\end{abstract}

\maketitle

\section{Introduction}

A Fock state is one of the most nonclassical states with a well-defined number of bosonic excitations such as photons, phonons, and magnons. It plays diverse and crucial roles in quantum information fields~\cite{RevModPhys1986Preparation}, including quantum communications~\cite{RevModPhys1994Caves}, quantum metrology~\cite{PhysRevLett1993Holland,Science2007TN,Natphot2011VG,Natphot2017SS}, and remote entanglement generation~\cite{PhysRevX2016AN}. Fock states are prominent candidates in open quantum systems to explore the environment-induced decoherence by measuring their decay time~\cite{PhysRevLett2008WH,PhysRevLett2008MB}. The superpositions of Fock states with particular components exhibit appealing features in fault-tolerant quantum information processing~\cite{QuantumST2021Jo}. For example, the superposed states with binomial coefficients for the Fock basis can be used in quantum error correction~\cite{PhysRevX2016Mi}. In addition, a high-precision measurement over weak classical signals can be achieved by using superposed Fock states~\cite{RevModPhys1980Caves} due to the squeezing effect induced by quantum interference between Fock components~\cite{PhysRevA1987W}. Generating desired Fock states and their superposition is therefore of significant interest to both the fundamental quantum physics and the applications in quantum technology.

Early proposals to generate the Fock state employed the resonator and the ancillary atom in the Jaynes-Cummings (JC) model~\cite{IOP2021Larson}, which are under full control in both initial states and coupling period. The atom serves as an intermediator between the resonator and the external driving field to pump quanta into the resonator by sequential half Rabi oscillations. The oscillation period depends on the current state of the resonator, by which the generation proposals demand elaborate external driving in the cavity QED setup~\cite{PhysRevLett1993Vo,PhysRevLett1996Law}. Based on such a stepwise method, Fock states and their superpositions were then experimentally generated in coplanar waveguide resonators for a microwave photon number $n\leq 6$ with a fidelity over $90\%$~\cite{Nature2008SHo,Nature2009SHo} and in a bulk acoustic-wave resonator for up to $n=7$ phonons with a fidelity over $26\%$~\cite{Nature2018chu}. A Fock state with $n\leq 3$ can be generated in the transmon-cavity platform by dispersively coupling the transmon qubit to the photonic resonator and using a two-photon stimulated Raman adiabatic passage~\cite{NatureCommu2017Pre}. Two-level atomic systems are not necessary in the step-by-step idea. A proposal in Ref.~\cite{PhysRevA2019Yan} was presented to generate the Fock state by the simultaneous adiabatic modulation over the driving frequency and intensity on a cavity with Kerr nonlinearity.

Beyond the methods based on multiple Rabi oscillations or equivalent transitions, Uria $et~al$. considered in a cavity QED setup, a two-level atom resonantly coupled to the target cavity field initially prepared in a coherent state~\cite{PhysRevLett2019Ur}. By numerical calculation, they found that after a preassigned period of free evolution, the cavity field evolves into a displaced Fock state of a large photon number with a fidelity over $70\%$. Recently, the experimental demonstration of a dissipation engineering over a three-dimensional microwave cavity was reported in Ref.~\cite{PhysRevLett2024Li}, which is based on a cascaded selective photon-addition operation assisted by an ancillary superconducting qubit and a readout resonator. The readout resonator was used to implement a quantum reservoir control for stabilizing a multiphoton Fock state of the cavity mode with a photon number up to $n=3$ with a fidelity over $77\%$.

Quantum measurement is found to be a promising candidate for state control tasks, such as quantum state purification~\cite{PhysRevLett2003Naka,PhysRevA2010Com,Wiseman_2006}, ground-state cooling of various quantum systems~\cite{PhysRevB2011LY,PhysRevB2020pU,PhysRevA2021Yan}, and nuclear spin polarization~\cite{PhysRevA2022Jin}. Also, it can be implemented in generating Fock states by elaborate parametric tuning or by nondemolition measurements. For example, a Fock state of a cavity field can be conditionally created by the postselection over the atomic level after a specified period of evolution in a cavity QED setup, in which the atom-field interaction is highly tunable to control the transition rate in a chosen atom-field subspace on resonance~\cite{PhysRevLett2001Fran}. An arbitrary superposed Fock state can be generated by the dispersive coupling between the cavity field and a qubit so that one could individually control the qubit transition rate associated with the corresponding qubit-field subspace by driving the qubit with external microwave signals containing multi frequency components~\cite{PhysRevLett.118.223604}. On performing quantum nondemolition measurement and feedback control over the field photon number to conditionally suppress the photon-number spread in Fock space~\cite{nature2007Gu}, a Fock state with a small photon number $n\leq 3$ was proposed in theory~\cite{PhysRevA2009Dot} and implemented in experiment~\cite{nature2011Sa}. Recently, quantum nondemolition measurement has been used in the preparation of a superposed Fock state in a circuit QED setup, which can be optimized by reinforcement learning~\cite{PhysRevA2024Per}.

With a JC Hamiltonian, we propose a protocol to conditionally generate a desired Fock state $|n\rangle$ and certain superposed Fock states, such as $(|0\rangle\pm|n\rangle)/\sqrt{2}$, of a target resonator that starts from the coherent state. The protocol is based on repeated projective measurements on the coupled qubit, which are separated with joint free evolution of the composite system lasting $\tau$. In the near-resonant case, $\tau$ is much shorter than that under the dispersive coupling. For the resonator, the evolution-and-measurement cycles induce a positive operator valued measure (POVM) described by $V_{ji}(\tau)=\langle j|e^{-iH\tau}|i\rangle$, where $|i\rangle$ and $|j\rangle$ represent the initial state of the ancillary qubit and the measurement outcome, respectively and $H$ denotes the full Hamiltonian of the composite system. Our protocol does not require any elaborate external driving or parametric tuning during the whole generation process. By appropriately choosing a POVM and the period $\tau$ for each evolution-measurement cycle, the selected Fock components can be gradually distinguished from the coherent state. By using a qutrit instead of a qubit and two nondegenerate resonators instead of a single one, our protocol can be extended to generate Bell-like states $(|00\rangle\pm|nn\rangle)/\sqrt{2}$. In the presence of decoherence, the outcome fidelity of our protocol is found to be limited by the trade-off between the projective measurements and the decay from the desired Fock components to their lower neighbors. In terms of the JC model and the projective measurement over the ancillary system, our work is inspired by previous proposals for generating nonclassical bosonic states, e.g., the Fock states~\cite{PhysRevA1996Har}, the classical mixture of Fock states~\cite{QuantumST2023Dela}, and the squeezed state in Ref.~\cite{PhysRevA2019Wei}. Nevertheless, our protocol is applicable to the generation of superposed states and extendable to multiple modes.

The rest of this work is organized as follows. In Sec.~\ref{model} we introduce the theoretical framework about the POVM based on the JC model and repeated projective measurements on the ancillary qubit. We find the reduction factor of the population distribution of the resonator, which is crucial to the ensuing state generation and is determined by the qubit-resonator coupling strength, their detuning, and the period of each evolution-and-measurement cycle. In Sec.~\ref{singlemodeFock} and Sec.~\ref{singlemodeFocksuper} we discuss the generation of various single-mode Fock states and their coherent superpositions, respectively. The protocol efficiency can be optimized by only modulating the evolution periods. In Sec.~\ref{doublemode} our protocol is extended to generate Bell-like state of multiple excitations in two nondegenerate resonator modes, which can be described by the two-dimensional population-reduction factors. We summarize the whole work in Sec.~\ref{conclusion}.

\section{Theoretical framework}\label{model}

\begin{figure}[htbp]
\centering
\includegraphics[width=0.9\linewidth]{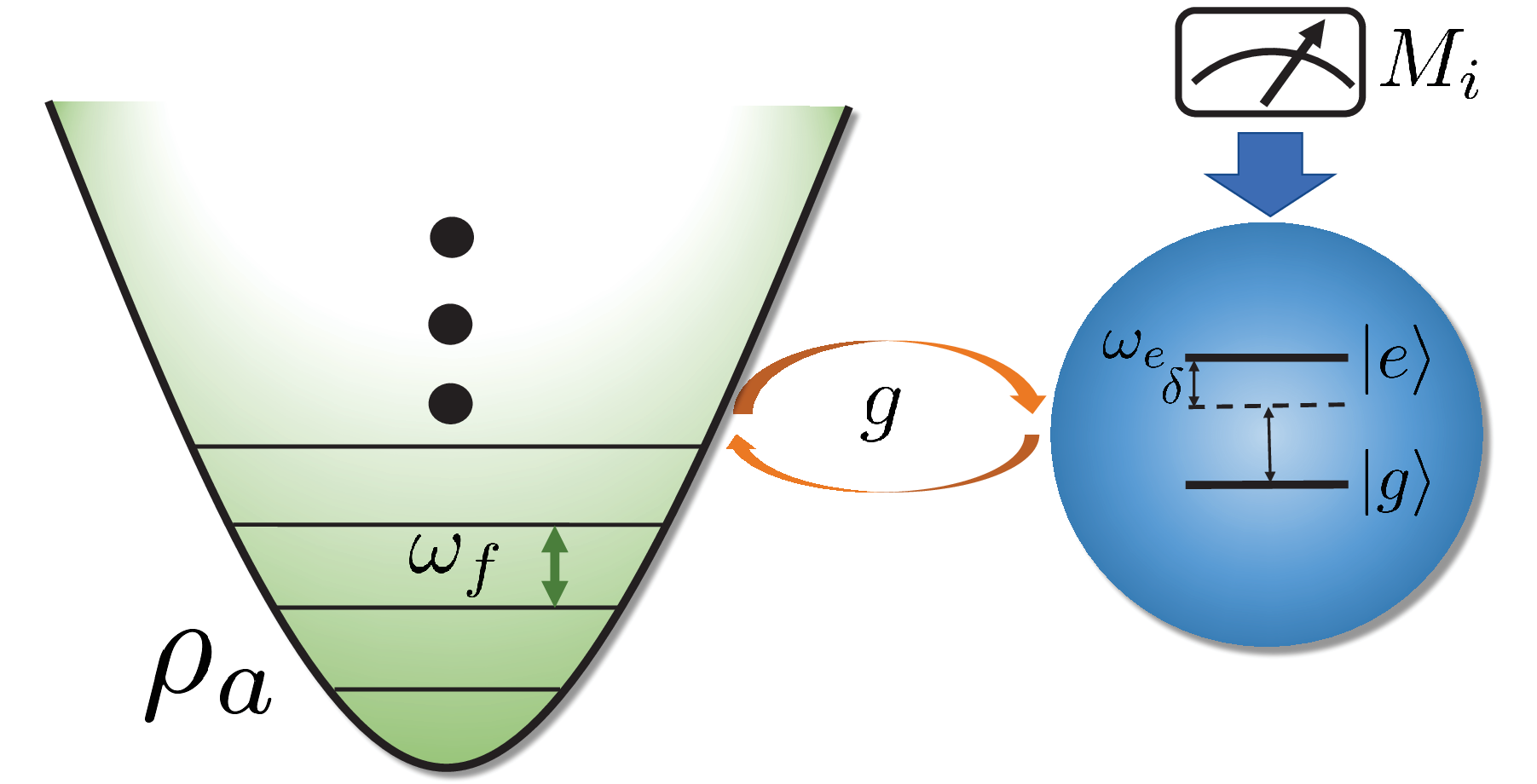}
\caption{Model diagram for our superposed Fock-state-generation protocol. A two-level system (qubit) serves as the ancillary system. The target resonator is coupled to the qubit through exchange interaction with strength $g$ and $\delta$ their detuning in frequency. The projective measurements $M_i=|i\rangle\langle i|$, $i=g,e$, are performed on the ancillary qubit.}\label{diagramJCM}
\end{figure}

To generate the desired Fock states or their superposition of a single bosonic mode, our protocol is based on a composite system consisting of a target resonator coupled to an ancillary qubit with the JC interaction~\cite{IOP2021Larson}, as shown in Fig.~\ref{diagramJCM}. The full Hamiltonian reads ($\hbar=1$)
\begin{equation}
  H=\omega_e|e\rangle\langle e|+\omega_fa^{\dagger}a+g\left(a^{\dagger}|g\rangle\langle e|+a|e\rangle\langle g|\right),
\end{equation}
where $\omega_e$ and $\omega_f$ are the level splitting of the ancillary qubit and the transition frequency of the target resonator, respectively, $a^\dagger$ ($a$) is the creation (annihilation) operator of the resonator, and $g$ denotes the qubit-resonator coupling strength. In the rotating frame with respect to $U_0=\exp[i\omega_f(|e\rangle\langle e|+a^{\dagger}a)t]$, the Hamiltonian becomes
\begin{align}\nonumber
     H'&=U_0HU_0^{\dagger}-iU_0\frac{\partial U_0^{\dagger}}{\partial t}\\ \label{Hprime} &=\delta|e\rangle\langle e|+g\left(a^{\dagger}|g\rangle\langle e|+a|e\rangle\langle g|\right),
\end{align}
where $\delta\equiv\omega_e-\omega_f$ denotes the detuning between the qubit and the resonator mode. Our theoretical framework is primarily established on the close-system Hamiltonian~(\ref{Hprime}) to distinguish the role played by projective measurement. The decoherence effect will be discussed in Secs.~\ref{singlemodeFock} and \ref{doublemode}.

Suppose that the ancillary qubit is prepared as one of its energy eigenstates $|i\rangle$, $i\in\{e,g\}$, and the resonator is initially at $\rho_a(0)$. After a period $\tau$ of joint free evolution governed by $U(\tau)=\exp(-iH'\tau)$, we perform a projective measurement on the qubit. If the measurement outcome is $|j\rangle$, $j\in\{e,g\}$, the density matrix of the resonator is then found to be
\begin{equation}
 \rho_a(\tau)=\frac{V_{ji}(\tau)\rho_a(0)V_{ji}^{\dagger}(\tau)}{{\rm Tr}\left[V_{ji}(\tau)
 \rho_a(0)V_{ji}^{\dagger}(\tau)\right]}=\frac{V_{ji}(\tau)\rho_a(0)V_{ji}^{\dagger}(\tau)}{P_j(1)},
\end{equation}
where $V_{ji}(\tau)=\langle j|e^{-iH'\tau}|i\rangle$ denotes an effective (nonunitary) evolution operator acting on the resonator and $P_j(1)$ represents the success probability of measuring the ancillary qubit on the state $|j\rangle$ after one evolution-measurement cycle. In our protocol, the cycle is repeated if $|j\rangle=|i\rangle$. Otherwise, the entire protocol is restarted from the very beginning.

The effective evolution operator $V_{ii}(\tau)$ then can be relabeled as $V_{i}(\tau)$ for simplicity, which is diagonal in the Fock-state basis $\{|k\rangle\}$~\cite{PhysRevB2011LY,PhysRevA2021Yan}, i.e.,
\begin{align}\label{Vi}
   V_i(\tau)=\sum_{k=0}^{\infty}\lambda_k^{(i)}(\tau)|k\rangle\langle k|.
\end{align}
Here the coefficient $\lambda^{(i)}_k(\tau)$ is
\begin{align}\label{coeff}
\lambda^{(i)}_k(\tau)&=e^{-i\delta t/2}\left[\cos\Omega^{(i)}_k\tau
+i\delta\dfrac{\sin\Omega^{(i)}_k\tau}{2\Omega^{(i)}_k}\right],
\end{align}
where the $k$-photon Rabi frequencies for $i=e,g$ are
\begin{align}
     \Omega_k^{(e)}&=\sqrt{\delta^2/4+(k+1)g^2},\\
     \Omega_k^{(g)}&=\sqrt{\delta^2/4+kg^2}.
\end{align}
respectively. Since $|\lambda^{(i)}_k(\tau)|\leq 1$, it works evidently as a reduction factor for the population on the Fock-state $|k\rangle$.

If the outcomes of each measurement are identical to the initial state of the qubit $|i\rangle$ and $\tau$ is invariant, the density matrix of the resonator for $N$ rounds of evolution and measurement is found to be
\begin{align}
\nonumber\rho_a(N\tau)&=\dfrac{V_{i}^N(\tau)\rho_a(0)V_{i}^{\dagger N}(\tau)}{\text{Tr}\left[V_{i}^N(\tau)\rho_a(0)V_{i}^{\dagger N}(\tau)\right]}\\
&=\mathcal{D}[\rho_a(N\tau)]+\mathcal{C}[\rho_a(N\tau)],\\ \label{Drhotau}
\mathcal{D}[\rho_a(N\tau)]&=\sum_{k=0}^{\infty}\frac{p_k|\lambda_k^{(i)}(\tau)|^{2N}}{P_i(N)}
|k\rangle\langle k|,\\ \label{Crhotau}
\mathcal{C}[\rho_a(N\tau)]&=\sum_{k\neq k'}\frac{C_{kk'}\left[\lambda_k^{(i)}(\tau)\lambda_{k'}^{(i)*}(\tau)\right]^N}{P_i(N)}|k\rangle\langle k'|,
\end{align}
where $\mathcal{D}[\rho_a]$ and $\mathcal{C}[\rho_a]$ denote the diagonal and off-diagonal parts of the density matrix, respectively. And $p_k\equiv\langle k|\rho_a(0)|k\rangle$ and $C_{kk'}\equiv\langle k|\rho_a(0)|k'\rangle$ are the populations and coherence of the initial density matrix of the resonator in the Fock-state basis $\{|k\rangle\}$. The success probability of finding the qubit at its initial state after $N$ measurements is found to be
\begin{align}\label{PiN}
    P_i(N)=\sum_{k=0}^{\infty}|\lambda^{(i)}_k(\tau)|^{2N}p_k.
\end{align}

According to Eqs.~(\ref{Drhotau}) and (\ref{Crhotau}), one can find that the variations of both the population $p_k$ and coherence magnitude $C_{kk'}$ are determined by the coefficient $\lambda^{(i)}_k(\tau)$. Note that $\lambda^{(i)}_k(\tau)$ is independent of the resonator state, and yet depends on the evolution period $\tau$, the resonator-qubit detuning $\delta$, and the resonator-qubit coupling strength $g$. By optimizing the evolution period $\tau$ for each round of evolution and measurement, one can construct a filter preserving the desired elements in the density matrix of the resonator with $|\lambda^{(i)}_k(\tau)|=1$ and in the same time eliminating the others with $|\lambda^{(i)}_k(\tau)|<1$. We stress that the filter function in our protocol can be used to construct the superposed Fock state without parametric modulation over system frequencies and coupling strength~\cite{LarsonJMO}.

\section{Fock-state and superposed Fock-state generation in single mode}

\subsection{Fock-state generation}\label{singlemodeFock}

Our protocol to generate an arbitrary Fock state $|n\rangle$ can be performed with a sequence of projective measurements $M_e=|e\rangle\langle e|$ on the excited state of ancillary qubit. In the beginning of the each round of evolution and measurement, the qubit starts from $|e\rangle$. The initial density matrix of the full system reads
\begin{align}
     \rho(0)=|e\rangle\langle e|\otimes|\alpha\rangle\langle\alpha|,\quad
     |\alpha\rangle=\sum_{k=0}^{\infty}\alpha_k|k\rangle,
\end{align}
where $\alpha_k=e^{-|\alpha|^2/2}\alpha^k/\sqrt{k!}$. On substituting Eq.~(\ref{coeff}) into Eq.~(\ref{Drhotau}), the diagonal part of the density matrix of the resonator after $N$ rounds of equally spaced evolution and measurement turns out to be
\begin{equation}\label{fockstate}
    \mathcal{D}[\rho_a(N\tau)]=\sum_{k=0}^{\infty}\dfrac{|\alpha_k|^2|\lambda_k^{(e)}(\tau)|^{2N}|k\rangle\langle k|}{P_e(N)}.
\end{equation}
The fidelity of the target Fock state $|n\rangle$ is found to be
\begin{equation}\label{FN}
    \mathcal{F}(N)=\langle n|\rho_a(N\tau)|n\rangle=\frac{|\alpha_n|^2|\lambda_n^{(e)}(\tau)|^{2N}}
    {\sum_k|\alpha_k|^2|\lambda_k^{(e)}(\tau)|^{2N}}.
\end{equation}

To protect the population on the state $|n\rangle$, the relevant population-reduction factor has to satisfy $|\lambda^{(e)}_n(\tau)|^2=1$. Due to Eq.~(\ref{coeff}), the evolution period of a single round $\tau$ can be set as
\begin{equation}\label{tau}
  \tau=l\tau_n^{(e)}=l\frac{\pi}{\Omega^{(e)}_n},
\end{equation}
where $l\subseteq\mathbb{N}^+$. If the irrelevant population-reduction factors $|\lambda^{(e)}_k(l\tau_k^{(e)})|^2$ for all the significantly occupied Fock states $|k\rangle$, $k\neq n$, are always less than unit, then the resonator state will gradually approach the desired $|n\rangle$ as the measurement number $N$ increases. The filter effect can be further amplified by the success probability $P_e(N)<1$ in the denominator of Eq.~(\ref{fockstate}). Therefore the density matrix of the resonator mode becomes
\begin{align}
    \rho_a(N\tau)\rightarrow\frac{|\alpha_n|^2}{P_e(N)}|n\rangle\langle n|=|n\rangle\langle n|
\end{align}
in the ideal large-$N$ limit. Also, it indicates that the success probability converges to the initial population of the target Fock state:
\begin{align}\label{PeN}
    P_e(N)\rightarrow|\alpha_n|^2=e^{-|\alpha|^2}\frac{|\alpha|^{2n}}{n!}.
\end{align}

\begin{figure}[htbp]
\includegraphics[width=0.9\linewidth]{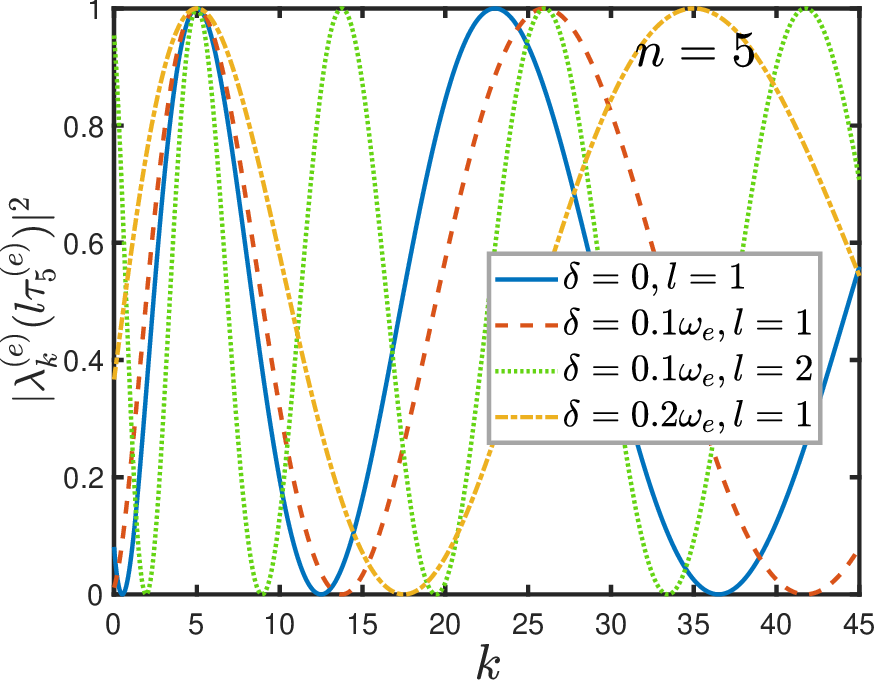}
\caption{Population-reduction factors for generating the Fock state $|5\rangle$ as a function of the Fock index $k$ after a single evolution-measurement cycle with various cycle periods $\tau$ determined by different detuning $\delta$ and integer $l$. Here $g=0.05\omega_e$.} \label{DETUNING}
\end{figure}

In Fig.~\ref{DETUNING} we present the population-reduction factors for generating the Fock state $|n=5\rangle$ as a function of the Fock index $k$ after one evolution-measurement cycle with optimized $\tau$. It is found with a fixed $l$ that a nonzero detuning $\delta$ gives rise to a smaller reduction rate for the neighboring states around $|5\rangle$ than a vanishing detuning. It will reduce our protocol's efficiency. Besides the target state $|n\rangle$, also the populations on the Fock states $|n_j^{(l)}\rangle$ remain invariant when
\begin{equation}\label{njl}
  n_j^{(l)}=\frac{j^2}{l^2}(n+1)+\left(\frac{j^2}{l^2}-1\right)\frac{\delta^2}{4g^2}-1
\end{equation}
with $j\subseteq \mathbb{N}^+$ due to Eqs.~(\ref{coeff}) and (\ref{tau}). Under the same $\delta$, the solution set with $l\neq1$ covers that with $l=1$ as shown by the red dashed line and green dotted line in Fig.~\ref{DETUNING}. To perform strategies for Fock-state generation with varying $\tau$ or $l$ in the sequence of measurements, we then focus on the intersection of the solution set with $l=1$ and that with $l\neq1$. Thus, in the following, $n_j^{(l)}$ can be briefly labeled as $n_j$ and it is readily found that $n=n_1$. More generally, a sufficiently large number of evolution-measurement cycles with $\tau$ in Eq.~(\ref{tau}) gives rise to a stabilized subspace $\mathcal{V}_n^{(e)}$ spanned by $\{\ket{n_j}|j\subseteq \mathbb{N}^+\}$, where both the populations and coherence are under protection. To enhance the final success probability of the target Fock state, the initial coherent state $\ket{\alpha}$ of the resonator can be prepared with $|\alpha|^2=n$ such that $|n\rangle$ is the maximally occupied Fock state and the populations on $|n_{j\geq2}\rangle$ are negligible. For example, when $n=n_1=5$ and $\delta=0$, the population on the next unwanted yet protected Fock-state $n_2=23$ is found to be $p_{n_2}\sim 10^{-5}$, about four orders lower than $p_{n_1}$.

\begin{figure}[htbp]
\includegraphics[width=0.9\linewidth]{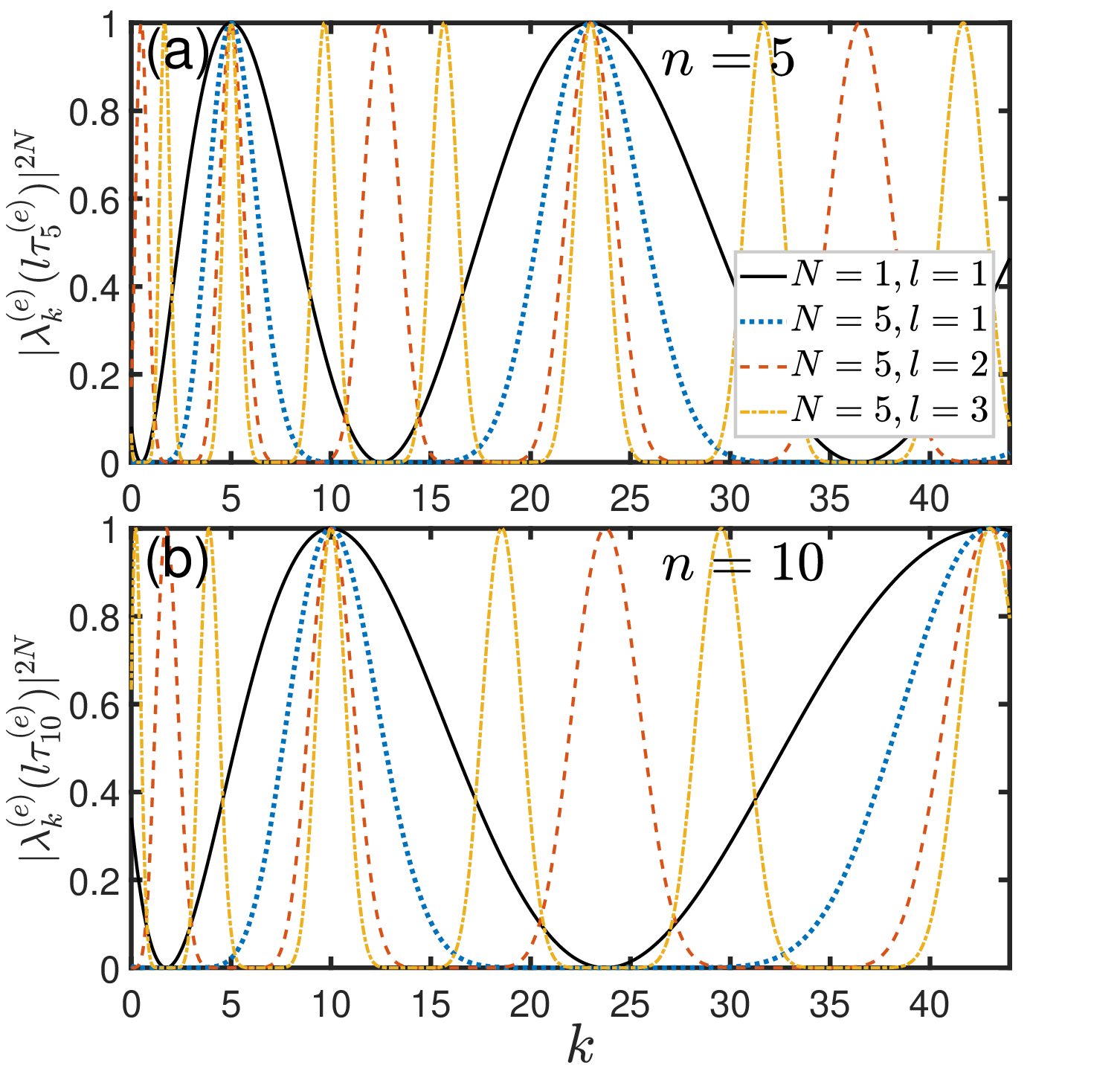}
\caption{Population-reduction factors as a function of the Fock index $k$ after $N$ evolution-measurement cycles with various evolution periods $\tau$ indicated by $l$. The target states are (a) $|n=5\rangle$ and (b) $|n=10\rangle$. The black solid lines indicate a single round with $\tau=\tau_n^{(e)}$ ($l=1$). The blue dotted lines, red dashed lines, and yellow dash-dotted lines indicate $N=5$ rounds with $\tau=\tau_n^{(e)}$, $\tau=2\tau_n^{(e)}$ ($l=2$), and $\tau=3\tau_n^{(e)}$ ($l=3$), respectively. Here $g=0.05\omega_e$ and $\delta=0$.} \label{lambdak}
\end{figure}

The efficiency for generating the target state is determined by the profile of the population-reduction factors $|\lambda^{(e)}_k(l\tau_n)|^{2N}$ in the Fock space as shown in Figs.~\ref{lambdak}(a) and \ref{lambdak}(b), where $n=5$ and $10$, respectively. These factors are evaluated for various periods $\tau$ of free-evolution (indicated by $l$) after $N$ rounds of evolution and measurement. The black solid lines in Fig.~\ref{lambdak} demonstrate the effect of a single projective measurement, confirming that the population-reduction factors of the irrelevant Fock states $k\neq n_{j\geq 1}$ satisfy $|\lambda_k^{(e)}(\tau)|^{2}<1$. Further, the comparison between the blue dotted lines and the black solid lines indicates that the populations in the neighboring regions of $|n_j\rangle$ can be significantly suppressed by a sufficient number of measurements.

Using multiples of $\tau$ (see the red dashed lines for $l=2$ and the yellow dashed-dot line for $l=3$), it is interesting to find that, on the one hand, they can intensify the population-reduction effect surrounding the target state $|n_1\rangle$ with a highly concentrated population-protection region; yet on the other hand, more protected (although not strictly) Fock states arise, e.g., see the high profiles between $|n_1=5\rangle$ and $|n_2=23\rangle$ in Fig.~\ref{lambdak}(a). Those results suggest a compromise or hybrid strategy labeled with $\mathcal{S}_l^{(q)}$ that filters out the populations distributed on the states other than $|n_j\rangle$ by using $\tau=\tau^{(e)}_{n}$ in the first $q$ rounds of evolution and measurement and then promotes the state-generation efficiency by using $\tau=l\tau^{(e)}_{n}$ with $l>1$ in the rest rounds. Often we take $l=2$ or $l=3$ since the advantage from an even larger $l$ is not significant. Accordingly, the uniform strategy with an invariant period $\tau=\tau^{(e)}_n$ can be labeled with $\mathcal{S}_1^{(\infty)}$.

\begin{figure}[htbp]
\centering
\includegraphics[width=0.9\linewidth]{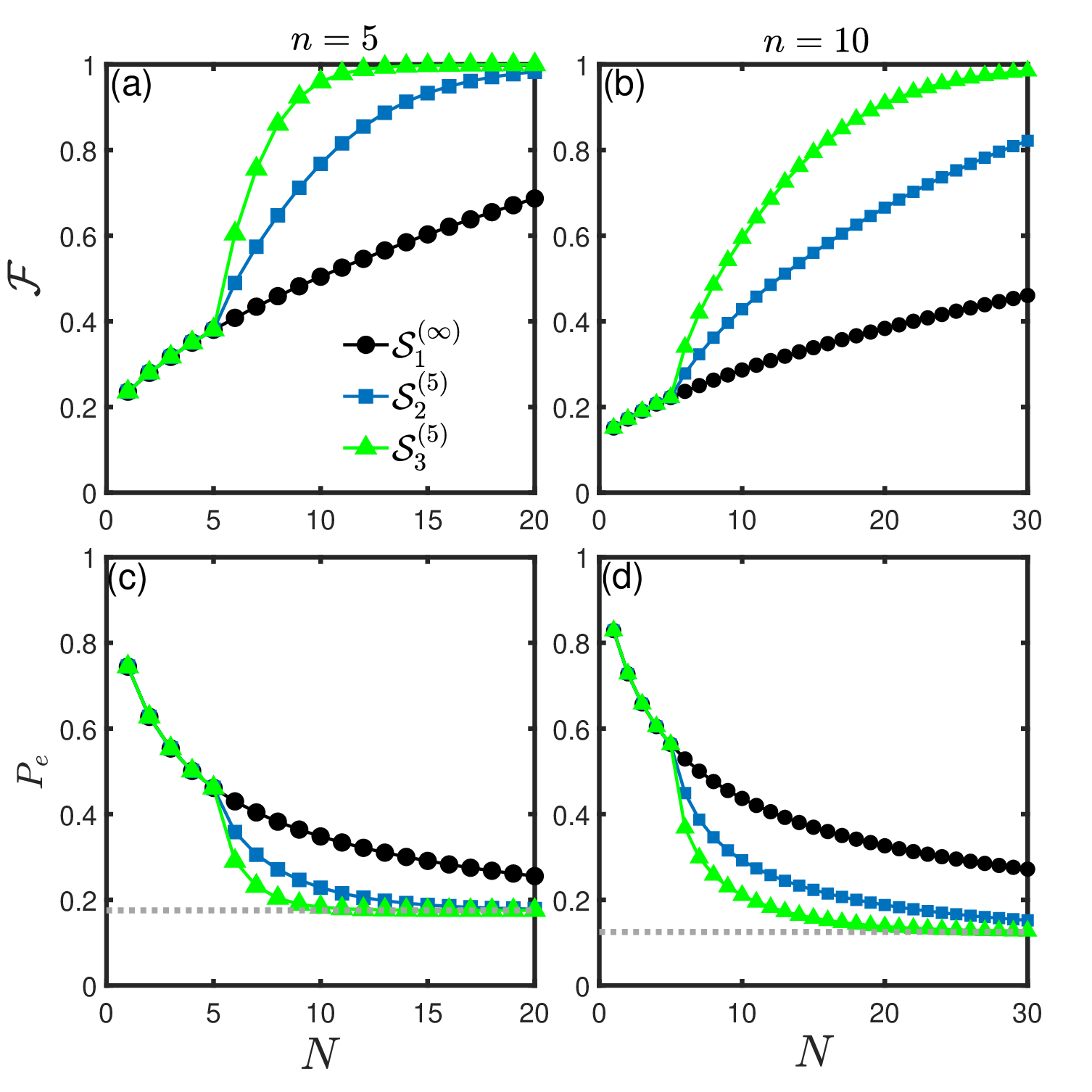}
\caption{Comparison of various strategies in Fock-state generation of (a) and (c) $|n=5\rangle$ and of (b) and (d) $|n=10\rangle$ as a function of the measurement number $N$. (a) and (b) Target-state fidelity. (c) and (d) Success probability, where the gray dotted lines indicate the initial population of $|n\rangle$. The system parameters are the same as in Fig.~\ref{lambdak}.}\label{fockgene}
\end{figure}

In Fig.~\ref{fockgene} we compare the efficiencies for generating the Fock state between the uniform strategy and the hybrid strategy $\mathcal{S}^{(5)}_l$ with $l=2,3$ in terms of the target-state fidelity $\mathcal{F}(N)$ and the success probability $P_e(N)$, whose definitions can be found in Eqs.~(\ref{FN}) and (\ref{PiN}), respectively. Within the current hybrid strategy, the first $q=5$ rounds of the evolution and measurement use $\tau=\tau_n^{(e)}$ and the remaining rounds use $\tau=l\tau_n^{(e)}$. It is found that the hybrid strategy overwhelms the uniform one in the state fidelity within a limited number of measurements. In addition, $l=3$ gives rise to the best performance. In Fig.~\ref{fockgene}(a) the fidelity $\mathcal{F}$ of the target Fock state $|5\rangle$ can reach $99.6\%$ within $15$ measurements when $l=3$. In contrast, $\mathcal{F}$ reaches only $68.6\%$ within $20$ measurements by the uniform strategy. In Fig.~\ref{fockgene}(b) it is found that a higher Fock state $|10\rangle$ could be obtained by merely $30$ measurements in the strategy $\mathcal{S}^{(5)}_3$ with a fidelity over $98.4\%$. A different strategy yields the same asymptotic value of the success probability, i.e., the population of $|n\rangle$ in the initial coherent state, which can be obtained by Eq.~(\ref{PeN}) with $|\alpha|^2=n$. In both Figs.~\ref{fockgene}(c) and \ref{fockgene}(d), a larger $l$ gives rise to a faster convergence.

\begin{figure}[htbp]
\centering
\includegraphics[width=0.9\linewidth]{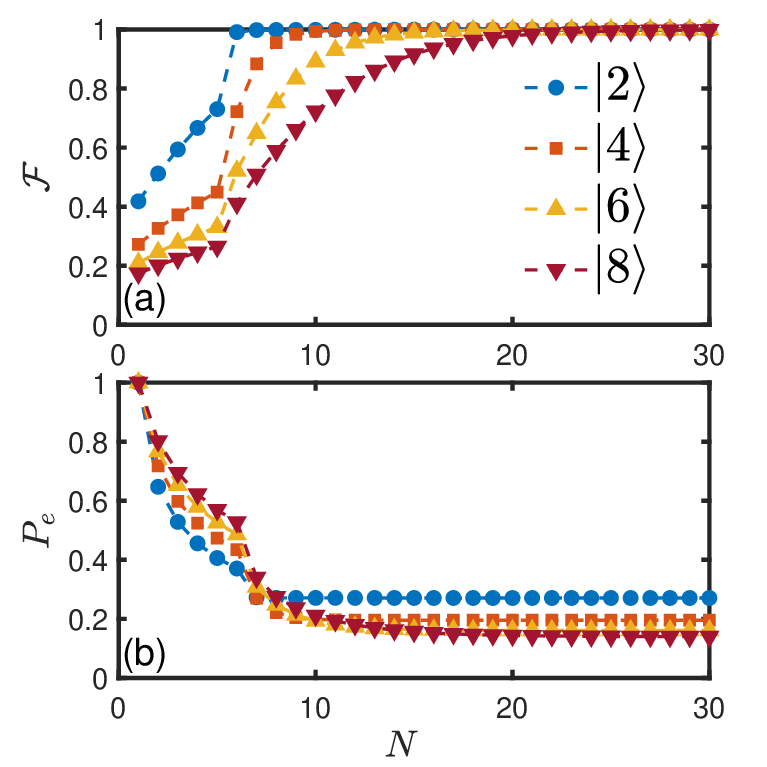}
\caption{Dependence of (a) fidelity and (b) success probability on the measurement number $N$ under the strategy $\mathcal{S}_3^{(5)}$ for various target state. The system parameters are the same as in Fig.~\ref{lambdak}.}\label{morefockgene}
\end{figure}

In Fig.~\ref{morefockgene} we generate different Fock states under the same hybrid strategy $\mathcal{S}_3^{(5)}$. The initial coherent states are accordingly set by $|\alpha|^2=n$. It is found in Fig.~\ref{morefockgene}(a) that more rounds of evolution-measurement cycles are required to prepare a higher Fock states. On the one hand, a larger $n$ indicates a broader population distribution in Fock space and a smaller overlap between the corresponding coherent state and the target state; and on the other hand, a larger $n$ leads to a smaller difference in Rabi frequencies associated with $|n\rangle$ and $|n\pm1\rangle$ and then more measurements are required to discriminate them. For example, it requires only $6$ rounds of evolution and measurement to enhance the fidelity of the target Fock state $|2\rangle$ to $99\%$, yet about $23$ rounds to prepare $|8\rangle$ with almost the same fidelity. Figure~\ref{morefockgene}(b) shows that the asymptotic behavior of the success probability is insensitive to $n$. It declines rapidly for the first few rounds and approaches the steady value after about $N=10$ rounds.
During the state-generation process, both target resonator and ancillary qubit cannot be completely isolated from the surrounding environment. The attainable fidelities of the target Fock states will be limited due to the coupling between the composite system and the environment (assumed to be at zero temperature). Under the Born-Markov approximation, the evolution of the composite system can be described by the Lindblad master equation:
\begin{equation}\label{master1}
  \begin{aligned}
    \dot{\rho}(t)=&-i[H', \rho(t)]+\gamma\mathcal{L}[|g\rangle\langle e|]\rho(t)\\ &+\kappa\mathcal{L}[a]\rho(t)+\gamma_{\phi}\mathcal{L}[|e\rangle\langle e|]\rho(t),
  \end{aligned}
\end{equation}
where $\gamma$ and $\kappa$ represent the decay rates of the ancillary qubit and the target resonator, respectively, and $\gamma_{\phi}$ is the dephasing rate of the qubit. For simplicity, we set $\gamma=\gamma_{\phi}=2\kappa$. The Lindblad superoperator $\mathcal{L}[\mathcal{O}]$ is defined as
\begin{equation}\label{Super}
\mathcal{L}[\mathcal{O}]\rho=\frac{1}{2}
(2\mathcal{O}\rho\mathcal{O}^{\dagger}-\mathcal{O}^{\dagger}\mathcal{O}\rho-\rho\mathcal{O}^{\dagger}\mathcal{O}),
\end{equation}

\begin{figure}[htbp]
\centering
\includegraphics[width=0.9\linewidth]{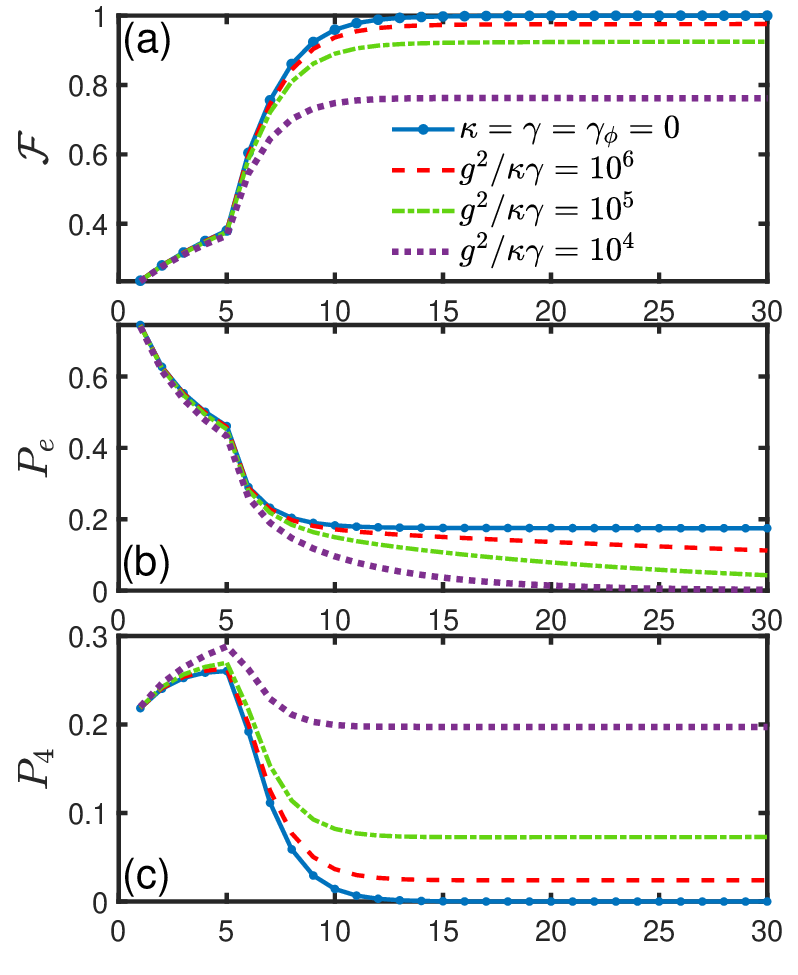}
\caption{Performance of our protocol in generating $|n=5\rangle$ under the strategy $\mathcal{S}_3^{(5)}$ in the presence of decoherence with various cooperativity $g^2/\kappa\gamma$. (a) Fidelity of the target state. (b) Success probability. (c) Population of the nearest-neighboring state $|4\rangle$. The system parameters are the same as in Fig.~\ref{lambdak}.}\label{fockgenedecay}
\end{figure}

In Fig.~\ref{fockgenedecay}(a), we present the fidelity for varying system cooperativity $g^2/\kappa\gamma$ for generating the target Fock state $|5\rangle$. In the presence of decoherence measured by the cooperativity $g^2/\kappa\gamma=10^6$, $10^5$, and $10^4$, the final fidelity of the target Fock state $|5\rangle$ can be stabilized at about $97\%$, $92\%$, and $76\%$, respectively, after a sufficient number of rounds of evaluation and measurement. It is the trade-off between the projective measurements and the decay from the desired Fock state to the lower states. In contrast to the fidelity, the success probability of measuring the ancillary qubit in its excited state $|e\rangle$ gradually decreases with $N$ due to the decoherence, as shown in Fig.~\ref{fockgenedecay}(b). To show the underlying reasons, we present the population dynamics of the Fock state $|4\rangle$ as the nearest-neighboring lower state of the target state in Fig.~\ref{fockgenedecay}(c). It is found that through the decay channel, $P_4$ increases in the first few rounds and then decreases with more measurements. However, the decoherence effect reduces the capability of our protocol to discriminate $|n-1\rangle$ from the target state $|n\rangle$, which results in a finite $P_4$ especially for a low system cooperativity.

\subsection{Superposed Fock-state generation}\label{singlemodeFocksuper}

Beyond the existing works for Fock state generation, our protocol can be used to prepare a superposed Fock state such as
\begin{align}\label{targetSF}
\ket{\psi_{\pm}}=c_0\ket{0}\pm c_n\ket{n},
\end{align}
with the desired $n$, $c_0$, and $c_n$, rather than a single Fock state~\cite{PhysRevA1996Har} or a mixture of Fock states~\cite{QuantumST2023Dela}. This target can be achieved by preparing the ancillary qubit in its ground state $|g\rangle$ and performing a sequence of projective measurements $M_g=|g\rangle\langle g|$ on the ancillary qubit. Due to Eqs.~(\ref{Vi}) and (\ref{coeff}), the effective time-evolution operator for the resonator under the resonant condition $\delta=0$ and an invariant $\tau$ is now written as,
\begin{align}
V_g(\tau)=\sum_{k=0}^{\infty}\cos\left(\sqrt{k}g\tau\right)|k\rangle\langle k|.
\end{align}
In this case, the population in the vacuum state $\ket{0}$ of the resonator is always under protection due to the decoupling of the state $\ket{g0}$ from the other composite states in the time evolution. Using Eqs.~(\ref{coeff}) and (\ref{Drhotau}), the diagonal part of the density matrix of the resonator after $N$ rounds of equally spaced evolution and measurement can be expressed as
\begin{align}\label{rhoNtauD2}
\mathcal{D}[\rho_a(N\tau)]&=\sum^{\infty}_{k=0}\frac{p_k\cos^{2N}\left(\sqrt{k}g\tau\right)}{P_g(N)}|k\rangle\langle k|,\\
P_g(N)&=\sum_{k=0}^{\infty}p_k\cos^{2N}\left(\sqrt{k}g\tau\right),
\end{align}
where $P_g(N)$ denotes the success probability of finding the qubit still at its ground state after $N$ measurements. Using Eq.~(\ref{Crhotau}), the off-diagonal part can be written as
\begin{align}\label{rhoNtauC2}
\mathcal{C}[\rho_a(N\tau)]&=\sum_{k\neq k'}\frac{C_{kk'}\left[\cos(\sqrt{k}g\tau)\cos(\sqrt{k'}g\tau)\right]^N}{P_g(N)}|k\rangle\langle k'|.
\end{align}
From Eqs.~(\ref{targetSF}), (\ref{rhoNtauD2}), and (\ref{rhoNtauC2}), the fidelity $\mathcal{F}_{\pm}$ about the target state $|\psi_{\pm}\rangle$ after $N$ equally-spaced evolution-and-measurement rounds can be obtained as
\begin{align}\nonumber
\mathcal{F}_{\pm}=&\bigg[|c_0|^2p_0+|c_n|^2p_n\cos^{2N}\left(\sqrt{n}g\tau\right) \\
&\pm2{\rm Re}(C_{n0}c_0c_n^*)\cos^N\left(\sqrt{n}g\tau\right)\bigg]/P_g(N). \label{fidelitypm}
\end{align}

\begin{figure}[htbp]
\centering
\includegraphics[width=0.9\linewidth]{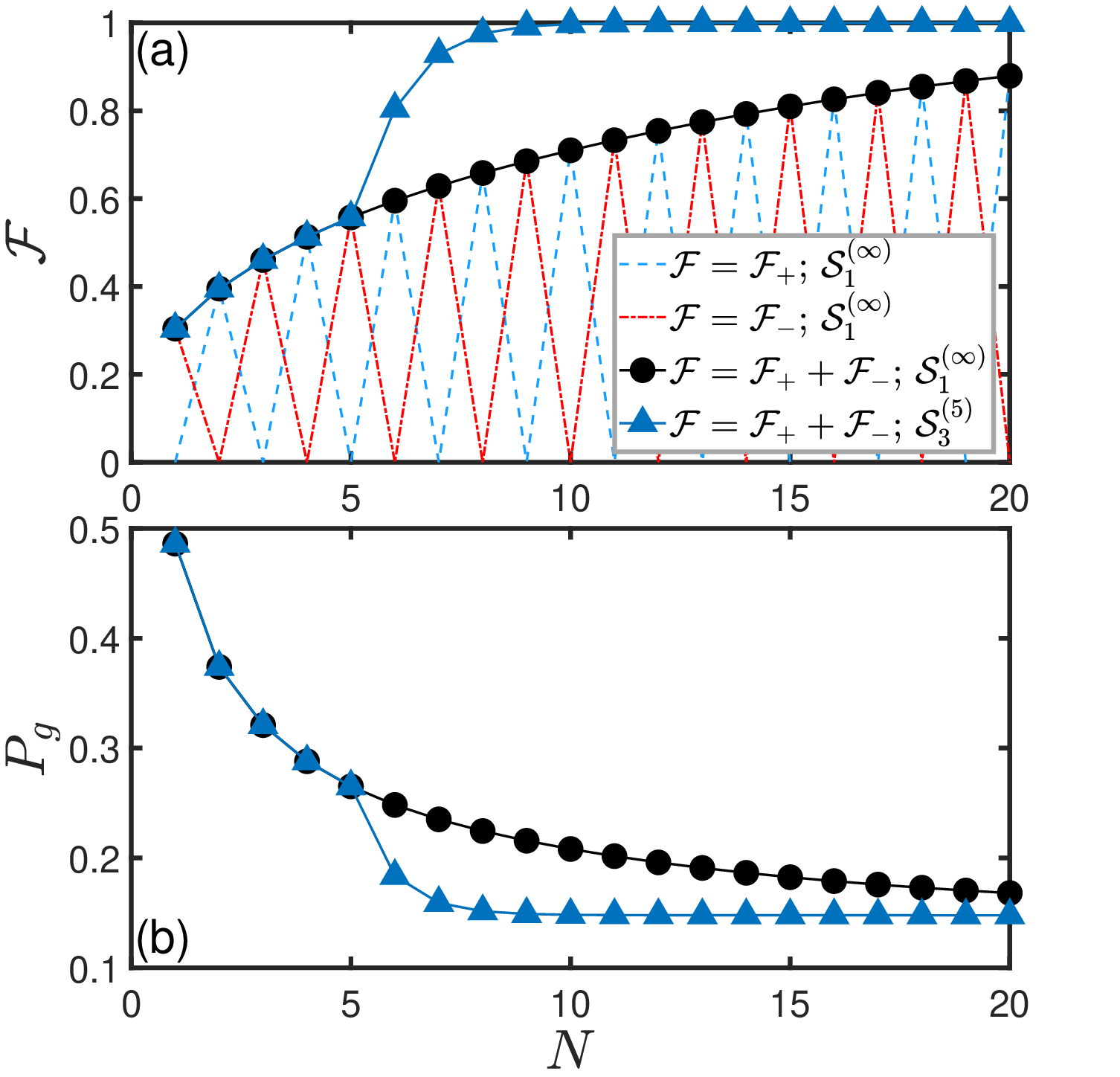}
\caption{(a) Fidelities $\mathcal{F}_{\pm}$ for the target superposed state $(|0\rangle\pm|5\rangle)/\sqrt{2}$ or their sum $\mathcal{F}_++\mathcal{F}_-$ as a function of $N$ under the uniform strategy $\mathcal{S}_1^{(\infty)}$ with $\tau=\tau_5^{(g)}$ and a hybrid strategy $\mathcal{S}_3^{(5)}$ that modifies the round period to be $\tau=3\tau_5^{(g)}$ after five rounds of evolution and measurement. (b) Success probability of finding the qubit in its ground state after $N$ measurements. The system parameters are the same as in Fig.~\ref{lambdak}.}\label{1modegenesuper1}
\end{figure}

Both populations in the ground state $|0\rangle$ and the Fock state $|n\rangle$ and their coherence should be under protection to generate the target superposed state $|\psi_{\pm}\rangle$ in Eq.~(\ref{targetSF}). To this end, the evolution periods for each round of evolution and measurement are chosen as ($\delta=0$)
\begin{equation}\label{taug}
  \tau=l\tau_n^{(g)}=l\frac{\pi}{\Omega^{(g)}_n}=l\frac{\pi}{g\sqrt{n}},
\end{equation}
where $l\in\mathbb{N}^+$. With $l=1$, a stabilized subspace under full protection emerges over multiple measurements. It reads
\begin{align}
\mathcal{V}^{(g)}_n={\rm span}\{|0\rangle, |n\rangle, |2^2n\rangle, |3^2n\rangle, ... \}.
\end{align}
The ratio of the probability amplitudes $c_0/c_n$ is determined by $\alpha_0/\alpha_n$. Then one can inversely find that the initial coherent state of the resonator is $|\alpha\rangle$ with $\alpha=(c_n\sqrt{n!}/c_0)^{1/n}$. Similar to the Fock-state generation in Sec.~\ref{singlemodeFock}, the initial populations on the Fock-state elements in the stabilized states $\{|j^2n\rangle|j\geqslant 2\}$ are negligible. Consequently, the diagonal and off-diagonal parts of the density matrix of the resonator in the large-$N$ limit read
\begin{align}\label{rhoNtauD}
\mathcal{D}[\rho_a(N\tau)]&\rightarrow e^{-|\alpha|^2}\left(|0\rangle\langle 0|+\frac{|c_n|^2}{|c_0|^2}|n\rangle\langle n|\right)\bigg /P_g(N),\\ \label{rhoNtauC}
\mathcal{C}[\rho_a(N\tau)]&\rightarrow e^{-|\alpha|^2}\left[\frac{c_n}{c_0}(-1)^{N}|n\rangle\langle 0|+{\rm H.c.}\right]\bigg /P_g(N),
\end{align}
respectively, where the success probably approaches
\begin{align}\label{PgN}
P_g(N)\rightarrow p_0+p_n=\exp\left[-\left(\frac{|c_n|}{|c_0|}\sqrt{n!}\right)^{2/n}\right]\bigg /|c_0|^2.
\end{align}
Eventually, the density matrix of the resonator results in
\begin{align}     \nonumber
\rho_a(N\tau)=&|c_0|^2|0\rangle\langle 0|+|c_n|^2|n\rangle\langle n|\\
     +&\left[(-1)^{N}c_nc_0^*|n\rangle\langle 0|+{\rm H.c.}\right]
\end{align}
by Eqs.~(\ref{rhoNtauD})--(\ref{PgN}).

\begin{figure}[htbp]
\centering
\includegraphics[width=0.9\linewidth]{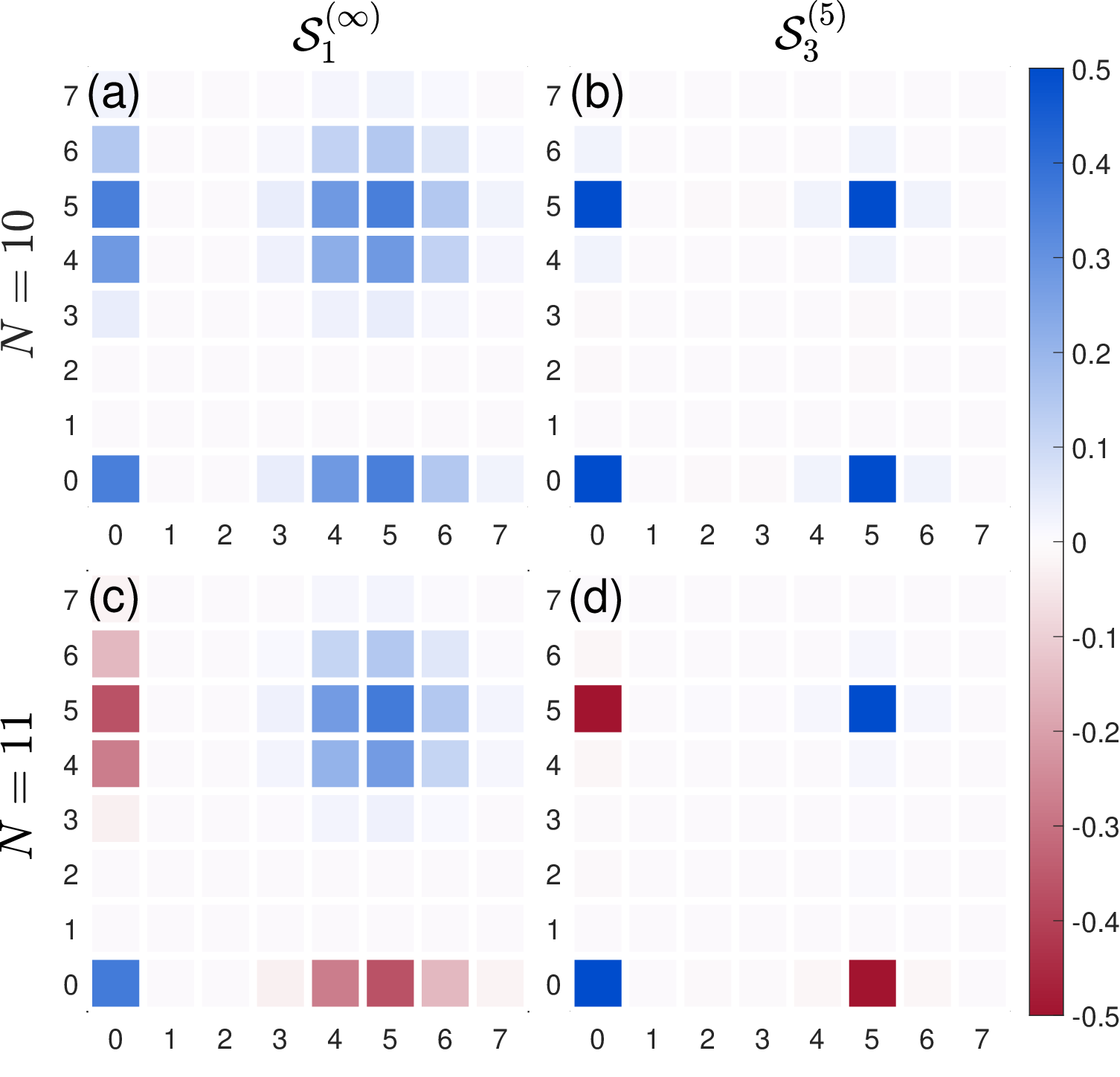}
\caption{Density matrix of the resonator after various rounds of evolution and measurement under (a) and (c) the uniform strategy or (b) and (d) the hybrid strategy $\mathcal{S}_3^{(5)}$ for (a) and (b) $N=10$ and (c) and (d) $N=11$. The target state is $(|0\rangle\pm|5\rangle)/\sqrt{2}$ and the other parameters are the same as in Fig.~\ref{lambdak}.}\label{densitymatri1}
\end{figure}

One can find that for $l=1$ (generally for an odd $l$), the parity of $N$ determining the output state is $|\psi_+\rangle$ or $|\psi_-\rangle$. In other word, after a sufficiently large number of measurements, the resonator state will oscillate between $|\psi_+\rangle$ and $|\psi_-\rangle$ with increasing fidelity. That alternate pattern can be observed from the blue dashed line and the red dash-dotted line in Fig.~\ref{1modegenesuper1}(a), which is consistent with Eq.~(\ref{fidelitypm}). One can therefore employ the sum $\mathcal{F}_++\mathcal{F}_-$ to characterize the performance of our protocol. At the end of each round of evolution and measurement, $\mathcal{F}_++\mathcal{F}_-=\mathcal{F}_+ (\mathcal{F}_-)$ for an even (odd) $N$. In Fig.~\ref{1modegenesuper1}(a), the fidelity $\mathcal{F}_++\mathcal{F}_-$ of the superposed state $(|0\rangle\pm|5\rangle)/\sqrt{2}$ is plotted for various strategies for an arrangement of $\tau$. In the same spirit as the Fock-state generation in Sec.~\ref{singlemodeFock}, the hybrid strategy displays a dramatic advantage over the uniform strategy. For example, the fidelity of the target state $|\psi_+\rangle$ reaches $99\%$ with only $N=10$ measurements under the strategy $\mathcal{S}_3^{(5)}$. In contrast, it is about $71\%$ if $\tau$ remains as $\tau_5^{(g)}$ with the same number of measurements. Figure~\ref{1modegenesuper1}(b) demonstrates the success probability of the qubit in its ground state $|g\rangle$ as a function of $N$. Again, the success probability using the hybrid strategy saturates at a faster rate than the uniform strategy.

The difference between uniform and hybrid strategies in the superposed-state generation can be unambiguously described by the density matrix of the resonator. For either $|\psi_+\rangle$ in Figs.~\ref{densitymatri1}(a) and \ref{densitymatri1}(b) or $|\psi_-\rangle$ in Figs.~\ref{densitymatri1}(c) and \ref{densitymatri1}(d), which can be respectively prepared by $N=10$ and $11$ rounds of measurements, respectively, it confirms that the hybrid strategy is superior to the uniform one.

\begin{figure}[htbp]
\centering
\includegraphics[width=0.9\linewidth]{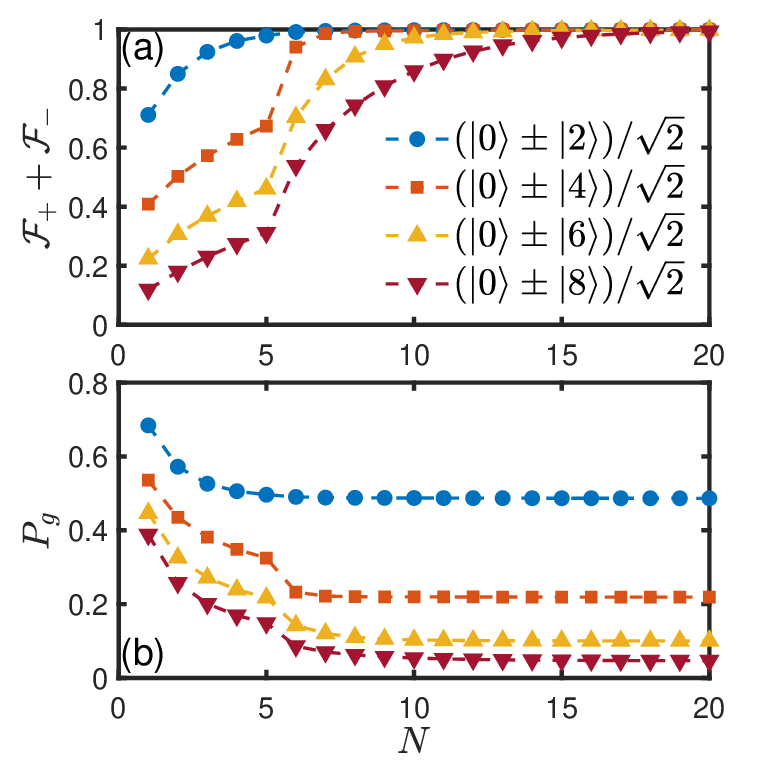}
\caption{Performance of our protocol in generating various superposed Fock states $(|0\rangle\pm|n\rangle)/\sqrt{2}$ with the hybrid strategy $\mathcal{S}_3^{(5)}$. (a) Sum of fidelity $\mathcal{F}_++\mathcal{F}_-$. (b) Success probability. The system parameters are the same as in Fig.~\ref{lambdak}. }\label{1modegenesuper2}
\end{figure}

Fidelities for more superposed Fock states are given in Fig.~\ref{1modegenesuper2}(a). Similar to the Fock-state generation in the closed-system scenario, more rounds of evolution and measurement are required for a larger $n$. It requires $N=6,~8,~14,$ and $20$ rounds of measurements to prepare the superposed states $(|0\rangle\pm|n\rangle)/\sqrt{2}$, with $n=2,~4,~6$, and $8$, respectively, with a fidelity over $99\%$. Note that the population distribution of any coherent state (the initial state of the resonator) determines that it is hard to find significant occupations in both the ground state $|0\rangle$ and Fock state $|n\rangle$ with a large $n$ at the same time. In contrast, if one focuses on the Fock-state generation, it is then always possible to find a coherent state with a peak occupation in the desired $|n\rangle$. As one can expect from Eq.~(\ref{PgN}), the success probability of $(|0\rangle\pm|8\rangle)/\sqrt{2}$ in Fig.~\ref{1modegenesuper2}(b) is much lower than that of $|8\rangle$ in Fig.~\ref{morefockgene}(b). In particular, the former is about $5\%$ and the latter is about $14\%$. As expected, the success probability for the superposed state also decreases with increasing $n$.

\section{Superposed Fock-state generation in double modes}\label{doublemode}

\begin{figure}[htbp]
\centering
\includegraphics[width=0.9\linewidth]{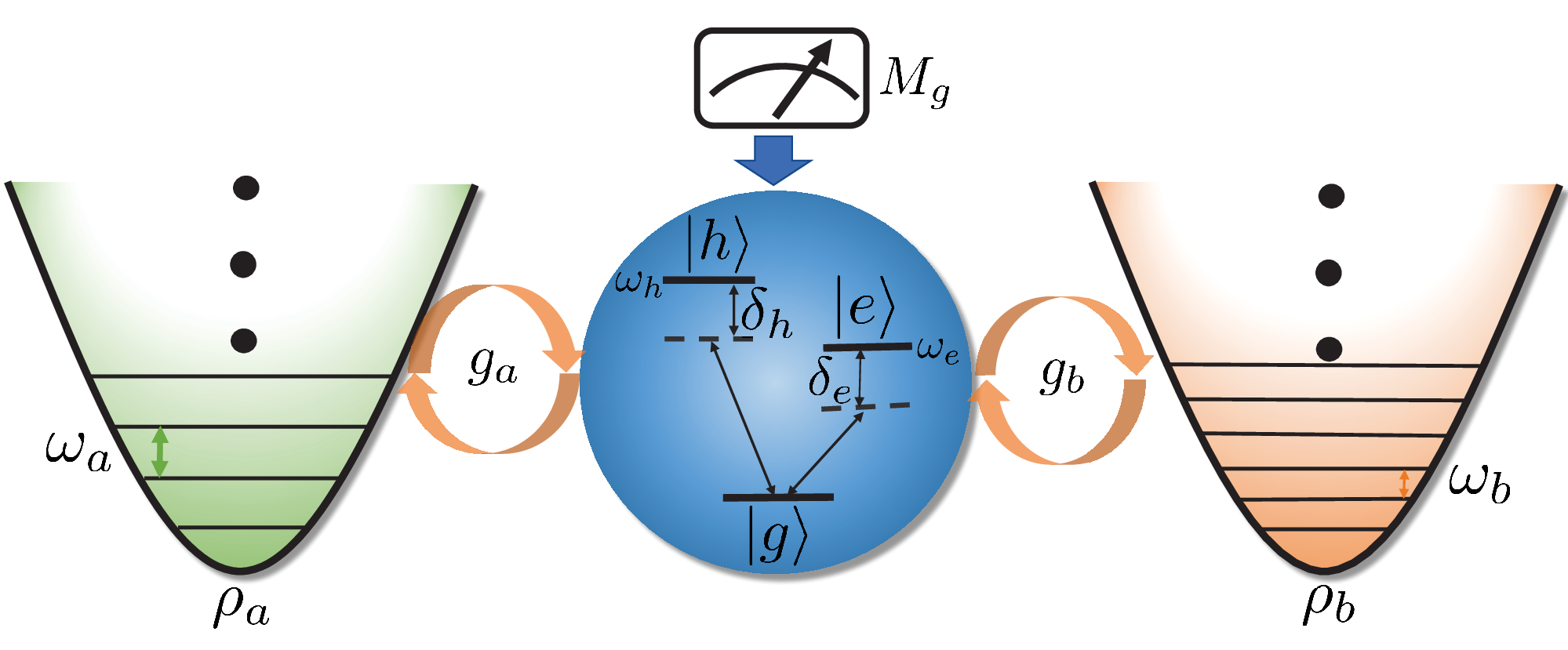}
\caption{Extended model diagram for our superposed Fock state generation protocol, in which a V-type three-level system (qutrit) is coupled to two nondegenerate resonator modes. The projective measurements $M_g=|g\rangle\langle g|$ performed on the ancillary qutrit can steer the double modes towards a multi excitation Bell state.}\label{diagram2mode}
\end{figure}

Replacing the ancillary qubit with a V-type qutrit, our protocol for generating a single-resonator Fock state can be extended to a double-resonator system with frequencies $\omega_a\neq\omega_b$, as shown in Fig.~\ref{diagram2mode}. The ground state, the middle state, and the highest-level state of the ancillary system are denoted by $|g\rangle$, $|e\rangle$, and $|h\rangle$, respectively. The resonator mode $a$ ($b$) is only coupled to the transition $|g\rangle\leftrightarrow|h\rangle$ ($|g\rangle\leftrightarrow|e\rangle$) with a coupling strength $g_a$ ($g_b$). In addition, $\delta_h=\omega_h-\omega_a$ and $\delta_e=\omega_e-\omega_b$ are the relevant detunings. In the rotating frame with respect to $U_0=\exp\{i[\omega_a(a^{\dagger}a+|h\rangle\langle h|)+\omega_b(b^{\dagger}b+|e\rangle\langle e|)]t\}$, the full Hamiltonian reads,
\begin{equation}
  \begin{aligned}
    H'&=\delta_h|h\rangle\langle h|+\delta_e|e\rangle\langle e| \\
    &+g_a(a^{\dagger}|g\rangle\langle h|+a|h\rangle\langle g|)+g_b(b^{\dagger}|g\rangle\langle e|+b|e\rangle\langle g|),
  \end{aligned}
\end{equation}
where $a^{\dagger}$ ($b^{\dagger}$) and $a$ ($b$) are the creation and annihilation operators for the resonator mode $a$ ($b$), respectively.

Our target is to generate a double-mode superposed Fock state $|\psi_{\pm}\rangle=c_{00}|0\rangle_a|0\rangle_b\pm c_{mn}|m\rangle_a|n\rangle_b$ (briefly labeled as $c_{00}|00\rangle\pm c_{mn}|mn\rangle$ in the following), which is essentially a general Bell state of multiple excitations. The qutrit is initially in the ground state $|g\rangle$. The two resonators $a$ and $b$ are prepared in coherent states $|\alpha\rangle$ and $|\beta\rangle$, respectively. To obtain the target state with desired amplitudes $c_{00}$ and $c_{mn}$, it is straightforward to find that $\alpha=(c_{mn}m!/c_{00})^{1/2m}$ and $\beta=(c_{mn}n!/c_{00})^{1/2n}$, in a similar way to generating a single-mode superposed state.

After $N$ rounds of evolution and measurement with a constant period $\tau$ and the measurement operator $M_g=|g\rangle\langle g|$, the state of the resonators reads
\begin{align}\label{rhoab}
\rho_{ab}(N\tau)=\frac{\Pi_g^N(\tau)|\alpha\rangle\langle\alpha|\otimes|\beta\rangle\langle\beta|\Pi_g^{\dagger N}(\tau)}{P_g(N)},
\end{align}
where $\Pi_g(\tau)=\langle g|e^{-iH't}|g\rangle$ denotes the effective time-evolution operator for the target resonators; $\Pi_g(\tau)$ is diagonal in the double-mode Fock basis $\{|kk'\rangle\equiv|k\rangle_a|k'\rangle_b\}$. In addition, $P_g(N)$ is the success probability that the qutrit remains at its ground state after $N$ measurements. To find a compact expression for the measurement operator, we set $\delta_g=\delta_e=\delta$. Then we have~\cite{PhysRevA2022Yan,SpringerPuri2001}
\begin{align}\label{Pig}
     \Pi_g(\tau)&=\sum_{k,k'}\lambda_{kk'}(\tau)|kk'\rangle\langle kk'|,
\end{align}
where the coefficients are
\begin{align}
\lambda_{kk'}(\tau)&=e^{-i(\delta\tau/2)}
\left(\cos\Omega_{kk'}\tau+i\frac{\delta}{2\Omega_{kk'}}\sin\Omega_{kk'}\tau\right),\\
\Omega_{kk'}&=\sqrt{\frac{\delta^2}{4}+g_a^2k+g_b^2k'}.
\end{align}

Using Eqs.~(\ref{rhoab}) and (\ref{Pig}), the diagonal part of the density matrix of the two resonator modes $a$ and $b$ in the basis $\{|kk'\rangle\}$ after $N$ rounds of equally spaced evolution and measurement is found to be
\begin{align}\label{Drhoab}
\mathcal{D}[\rho_{ab}(N\tau)]=\sum_{k,k'}\frac{|\lambda_{kk'}(\tau)|^{2N}p_{kk'}}{P_g(N)}|kk'\rangle\langle kk'|,
\end{align}
where $p_{kk'}\equiv\langle kk'|\rho_a(0)\otimes\rho_b(0)|kk'\rangle$ is their initial population and the success probability turns out to be
\begin{align}
P_g(N)=\sum_{k,k'}|\lambda_{kk'}(\tau)|^{2N}p_{kk'}.
\end{align}
The population-reduction factor of the double-mode Fock states $|kk'\rangle$ is determined by the modulus square of the coefficients $|\lambda_{kk'}(\tau)|^{2}$. The matrix elements of the off-diagonal part reads
\begin{align}\label{Crhoab}
\bra{kk'}\mathcal{C}[\rho_a(N\tau)]\ket{mm'}=&\frac{C_{kk',mm'}\left[\lambda_{kk'}(\tau)\lambda^*_{mm'}(\tau)\right]^N}{P_g(N)},
\end{align}
where $C_{kk',mm'}\equiv\langle kk'|\rho_a(0)\otimes\rho_b(0)|mm'\rangle$ represents the coherence of the initial state.

Under the resonant condition $\delta=0$, it turns out that $\lambda_{00}(\tau)=1$, irrespective of the evolution period $\tau$. In this case, $|g00\rangle$ is decoupled from all the other states of the composite system. From Eqs.~(\ref{Drhoab}) and (\ref{Crhoab}), the fidelity of the target states $|\psi_{\pm}\rangle$ is found to be
\begin{equation}\label{fidelitypm2}
  \begin{aligned}
     \mathcal{F}_{\pm}=&\Big[|c_{00}|^2p_{00}+|c_{mn}|^2p_{mn}\cos^{2N}\left(\Omega_{mn}\tau\right) \\
     \pm &2{\rm Re}\left(c_{mn}^*c_{00}C_{mn,00}\right)\cos^N\left(\Omega_{mn}\tau\right)\Big]/P_g(N).
  \end{aligned}
\end{equation}

To generate $|\psi_{\pm}\rangle=c_{00}|00\rangle\pm c_{mn}|mn\rangle$, or equivalently to preserve both the population on $|mn\rangle$ and the coherence between $|00\rangle$ and $|mn\rangle$, the evolution period $\tau$ for each round of evolution and measurement can be chosen as $\tau=l\tau_{mn}=l\pi/\Omega_{mn}$ with $l\in\mathbb{N}^+$ to guarantee the condition $|\lambda_{mn}(\tau)|=1$. When $l$ is odd, the parity of $N$ will determine whether or not a $\pi$ phase presents in the coherence $\langle00|\mathcal{C}[\rho_a(N\tau)]|mn\rangle$, i.e., the outcome state is $|\psi_+\rangle$ or $|\psi_-\rangle$. Further, similar to the preceding situations, the asymptotic success probability is equivalent to the overlap between the target states and the initial state of the resonator system. It is found to be
\begin{align}\label{Pg2mode}\nonumber
&P_g(N)\rightarrow|\alpha_0\beta_0|^2+|\alpha_m\beta_n|^2\\
=&\exp\left[-\left(\frac{|c_{mn}|}{|c_{00}|}n!\right)^{1/n}-\left(\frac{|c_{mn}|}{|c_{00}|}m!\right)^{1/m}\right]
/|c_{00}|^2.
\end{align}

\begin{figure}[htbp]
\centering
\includegraphics[width=0.9\linewidth]{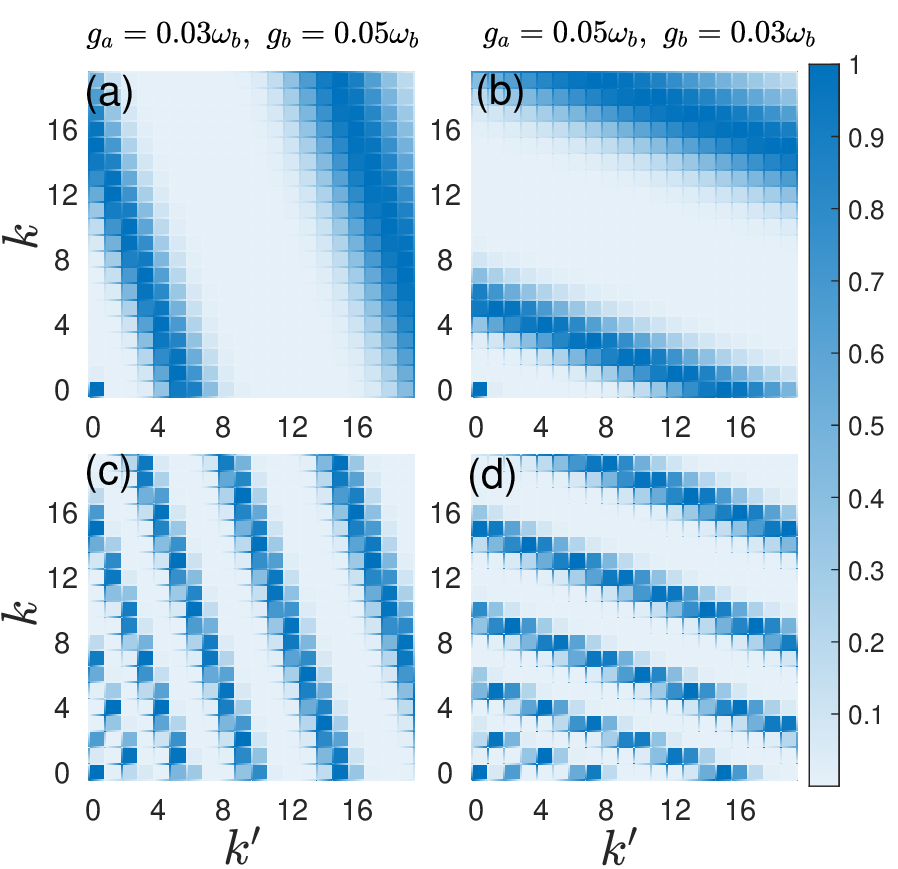}
\caption{Two-dimensional population-reduction factors $|\lambda_{kk'}(\tau)|^{2N}$ with $N=5$ in the double-mode Fock-state basis $\{|kk'\rangle\}$ for various coupling strengths and measurement intervals, with (a) and (c), $g_a=0.03\omega_b$ and $g_b=0.05\omega_b$ and (b) and (d) $g_a=0.05\omega_b$ and $g_b=0.03\omega_b$ and (a) and (b) $\tau=\tau_{44}$ and (c) and (d) $\tau=3\tau_{44}$.}\label{lambdakkp}
\end{figure}
  
Different from the single-mode Fock-state generation in Sec.~\ref{singlemodeFock}, we have an extra degree of freedom in modulating the efficiency for generation of a double-mode state, since the double-mode population reduction factor $|\lambda_{kk'}(\tau)|^{2N}$ actually has a two-dimensional profile, as shown in Fig.~\ref{lambdakkp}. The target state is assumed to be $|\psi_{\pm}\rangle=c_{00}|00\rangle\pm c_{44}|44\rangle$. However, many states other than the desired Fock states $|00\rangle$ and $|44\rangle$ also present with a unit or close-to-unit population-reduction factor. One can find in Fig.~\ref{lambdakkp} that these unwanted states existing in the dark-blue areas are distributed along the lines with a slope $-g_b/g_a$ passing through the points $(4j^2, 4j^2)$, $j\in\mathbb{N}^+$. Note that Figs.~\ref{lambdakkp}(a) and \ref{lambdakkp}(b), we have $g_a/g_b=3/5$ and $5/3$, respectively. The two profiles are dramatically different from each other while their crossing points are the desired Fock states $|00\rangle$ and $|44\rangle$. A solution to reduce the populations of the unwanted states is then to alter the coupling-strength ratio in the whole sequence of evolution-and-measurement rounds.

\begin{figure}[htbp]
\centering
\includegraphics[width=0.9\linewidth]{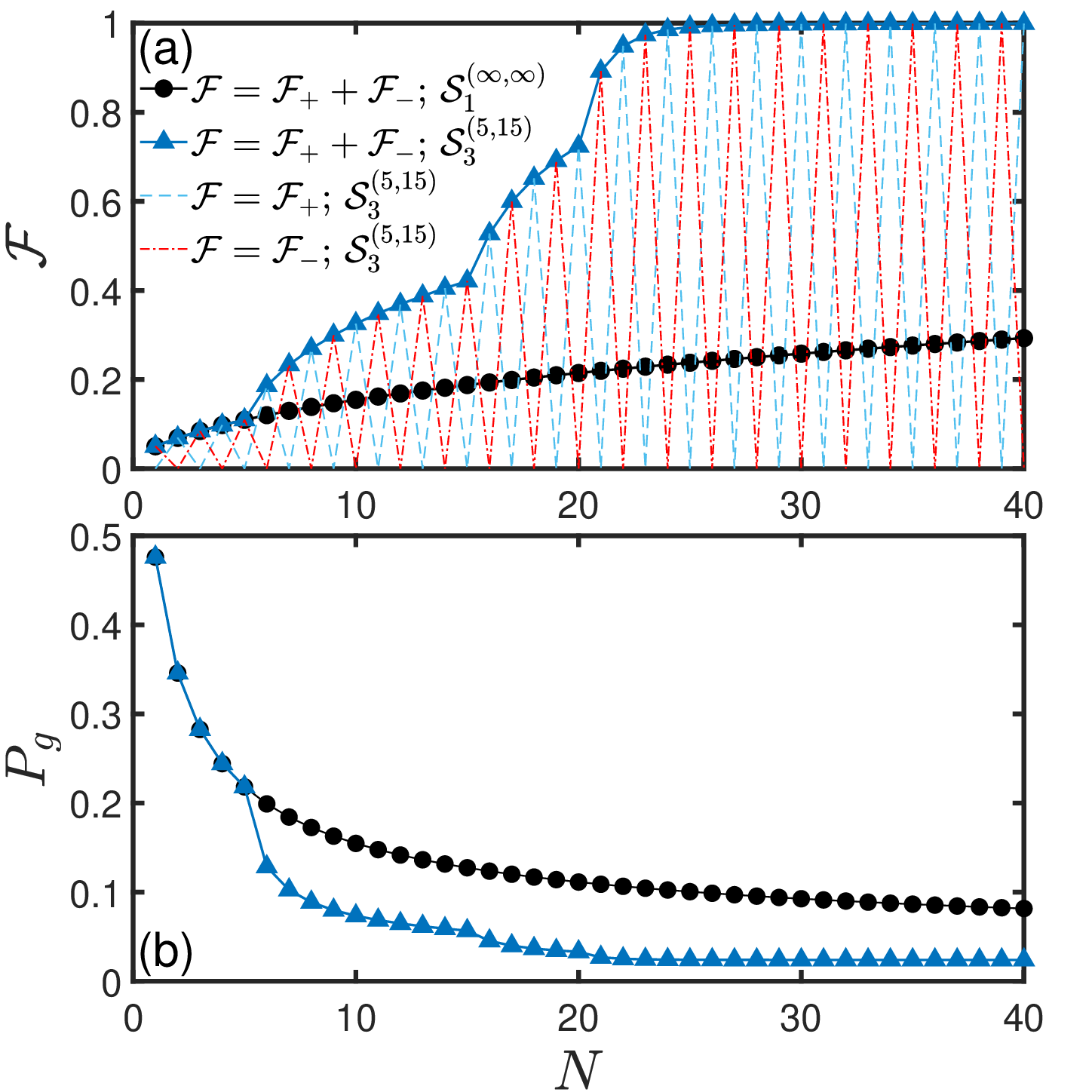}
\caption{(a) Fidelities of the target state $(|00\rangle\pm|44\rangle)/\sqrt{2}$ after $N$ rounds of evolution and measurement under the uniform strategy with fixed $\tau=\tau_{44}$, $g_a=0.05\omega_b$, and $g_b=0.03\omega_b$ or the hybrid strategy $\mathcal{S}_3^{(5,15)}$, in which the coupling strengths $g_a=0.05\omega_b$ and $g_b=0.03\omega_b$ are modified to be $g_a=0.03\omega_b$ and $g_b=0.05\omega_b$ after $L=15$ rounds of measurements. (b) Success probability of finding the qutrit in its ground state after $N$ rounds.}\label{2modegenesuper}
\end{figure}

Similar to the single-mode case (see Fig.~\ref{lambdak}), a longer evolution period $\tau$, i.e., a larger $l$, results in a more concentrated population-protection region in Fock space, indicating a higher generation efficiency. However, it also gives rise to more protected regions. These two features can be identified in Figs.~\ref{lambdakkp}(c) and \ref{lambdakkp}(d) by comparing to Figs.~\ref{lambdakkp}(a) and \ref{lambdakkp}(b), respectively. Thus, in regard to the two-dimensional profile of the population-reduction ratio, the hybrid strategy for the single-mode case is updated to have various ratios of the coupling strengths $g_a/g_b$ and various periods $\tau$. It can be relabeled as $\mathcal{S}^{(q,L)}_l$, $L>q$. In the following evaluation, the coupling ratio $g_a/g_b$ is modified from $5/3$ to $3/5$ after $L$ rounds of measurements; the round period is $\tau=\tau_{mn}$ for $1\leq N\leq q$ and $L+1\leq N\leq L+q$ and becomes $\tau=l\tau_{mn}$ for the remaining rounds. The uniform strategy with invariant $\tau=\tau_{mn}$ and $g_a/g_b=5/3$ can be labeled as $S_1^{(\infty,\infty)}$. Note that the strategies apply to arbitrary $g_a/g_b$.

In Fig.~\ref{2modegenesuper}(a) we present the fidelity of the target state $(|00\rangle\pm|44\rangle)/\sqrt{2}$ as a function of the measurement number $N$ under either uniform strategy or hybrid strategy $\mathcal{S}^{(5,15)}_{3}$. Here $\mathcal{F}_+$ and $\mathcal{F}_-$ manifest the same alternate pattern for $N$ of different parity as in Fig.~\ref{1modegenesuper1}(a). With respect to the fidelity sum $\mathcal{F}_++\mathcal{F}_-$, the hybrid strategy $\mathcal{S}^{(5,15)}_3$ demonstrates a significant advantage over the uniform strategy. With $30$ rounds of measurements, the former can prepare the Bell-like state with a fidelity over $99\%$, while the latter attains only $25\%$ in fidelity. Figure~\ref{2modegenesuper}(b) indicates the cost of the target-state preparation, in which the success probability is stabilized around about $3\%$ after about $25$ rounds of evolution and measurement.

\begin{figure}[htbp]
\centering
\includegraphics[width=0.9\linewidth]{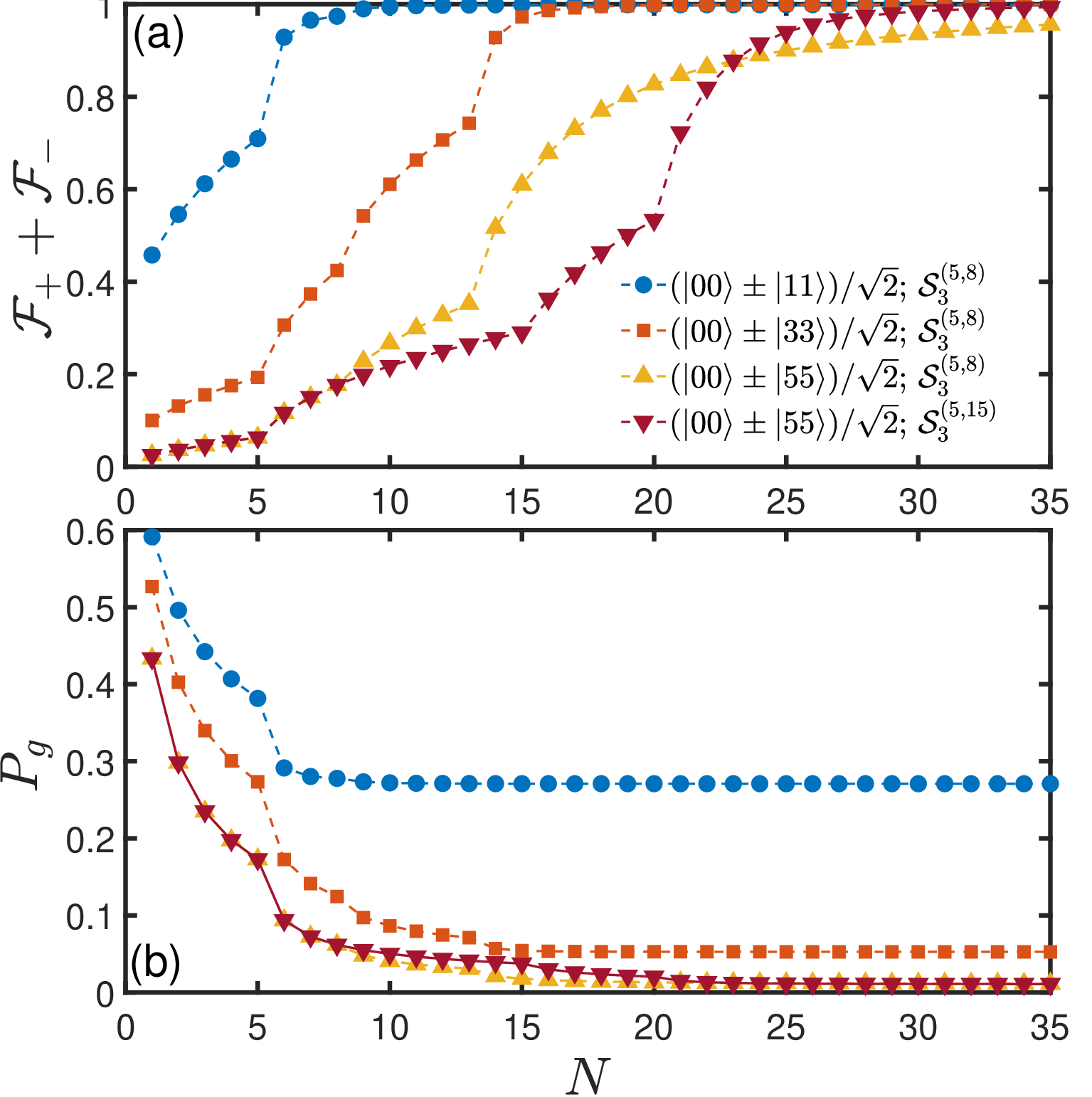}
\caption{Performance of our protocol in generating various Bell-like states $(|00\rangle\pm|nn\rangle)/\sqrt{2}$ under various hybrid strategies. (a) Fidelity sum $\mathcal{F}_++\mathcal{F}_-$. (b) Success probability. The system parameters are the same as in Fig.~\ref{2modegenesuper}.}\label{2modegenesuper2}
\end{figure}

In Fig.~\ref{2modegenesuper2}, we show the performance of our protocol in generating Bell-like states $(|00\rangle\pm|nn\rangle)/\sqrt{2}$ with $n=1$, $3$, and $5$ under various hybrid strategies. Under the same strategy $S_3^{(5,8)}$, the fidelities of the target states with $n=1$ and $3$ reach $99\%$ after $N=9$ and $16$ rounds of measurements. Their success probabilities approach $27\%$ and $5\%$, respectively. It is evident that it is harder to prepare a Bell-like state with a larger $n$. The comparison between different strategies for the same target state $(|00\rangle\pm|55\rangle)/\sqrt{2}$ indicates that it is nontrivial to select a proper strategy in state generation. During the first $L$ rounds with a constant ratio $g_a/g_b$, the unwanted yet protected populations should be sufficiently suppressed. Otherwise, the performance as a whole will become less optimized. That could explain why the final fidelity under $S_3^{(5,8)}$ is lower than that under $S_3^{(5,15)}$. In the latter case, one can prepare $(|00\rangle\pm|55\rangle)/\sqrt{2}$ in $N=33$ rounds of measurements with a fidelity over $99\%$.

\begin{figure}[htbp]
\includegraphics[width=0.9\linewidth]{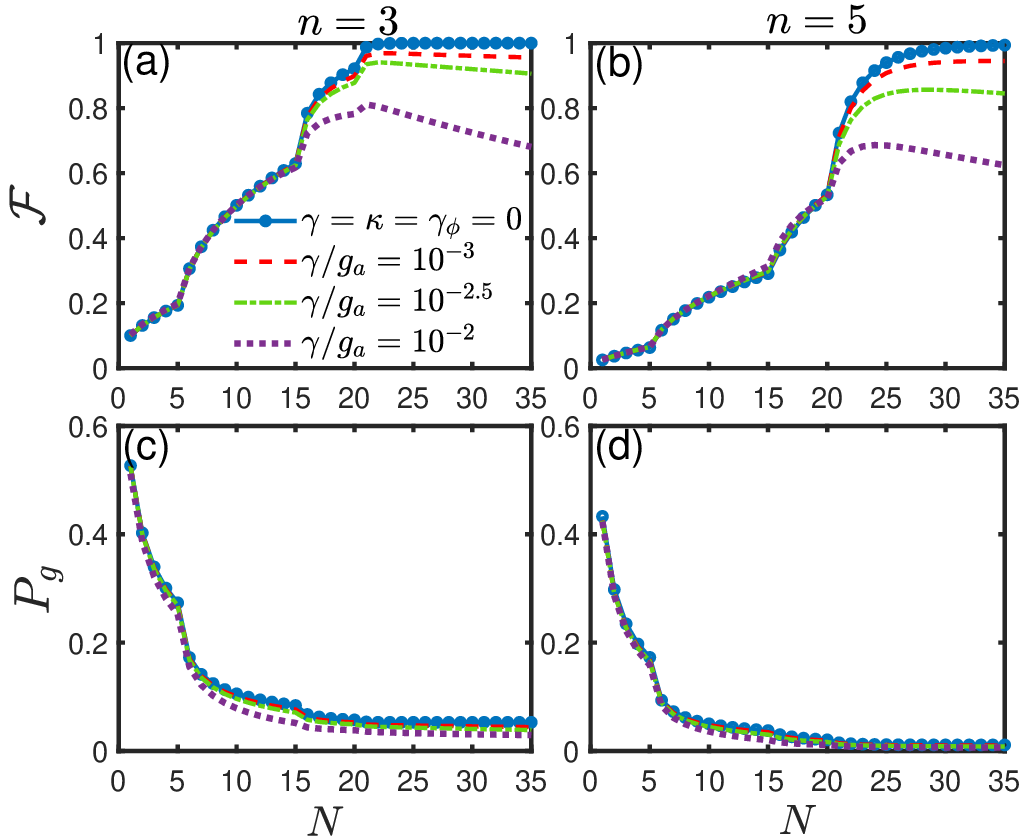}
\caption{Performance of our protocol for generating $(|00\rangle\pm|nn\rangle)/\sqrt{2}$ after $N$ rounds of evolution and measurement with various decoherence rate under the hybrid strategy $\mathcal{S}_3^{(5,15)}$ for $g_a=0.05\omega_e$, $g_b=0.03\omega_e$, and (a) and (c) $n=3$ and (b) and (d) $n=5$. (a) and (b): Fidelity of the target Bell-like states. (c) and (d): Success probability.} \label{2modegenesuperdecay}
\end{figure}

By taking all the relevant dissipation and dephasing channels into account, the Lindblad master equation for the two-resonator-qutrit system can be written as 
\begin{equation}\label{master2}
\begin{aligned}
 \dot{\rho}(t)=&-i[H', \rho{(t)}]+\gamma_{ge}\mathcal{L}[|g\rangle\langle e|]\rho(t)+\gamma_{eh}\mathcal{L}[|e\rangle\langle h|]\rho(t)\\ &+\gamma_{gh}\mathcal{L}[|g\rangle\langle h|]\rho(t)+\kappa_a\mathcal{L}[a]\rho(t)+\kappa_b\mathcal{L}[b]\rho(t)\\ &+\gamma_{e,\phi}\mathcal{L}[|e\rangle\langle e|]\rho(t)+\gamma_{h,\phi}\mathcal{L}[|h\rangle\langle h|]\rho(t),
\end{aligned}
\end{equation}
where $\kappa_i$ with $i=a,b$ represents the decay rate of resonator mode $i$, $\gamma_{ij}$ with $i,j=g,e,h$ represents the dissipation rate from the qutrit level $|j\rangle$ to $|i\rangle$, and $\gamma_{i,\phi}$ with $i=e,h$ represents the dephasing rate of the qutrit level $|i\rangle$. Here we set $\kappa=\kappa_a=\kappa_b$, $\gamma=\gamma_{ge}=\gamma_{eh}=\gamma_{gh}=\gamma_{e,\phi}=\gamma_{h,\phi}$, and $\kappa=0.5\gamma$.

In Figs.~\ref{2modegenesuperdecay}(a) and \ref{2modegenesuperdecay}(b) we present the fidelities of a two-mode Bell-like state $(|00\rangle\pm|nn\rangle)/\sqrt{2}$ with $n=3$ and $5$, respectively, in the presence of decoherence. Here the fidelity $\mathcal{F}=\mathcal{F}_+$ when $N$ is even and $\mathcal{F}=\mathcal{F}_-$ when $N$ is odd. It is found that the Bell-like state can be approximately obtained only for a low decoherence rate $\gamma/g_a=10^{-3}$. On performing more and more rounds of measurements, an optimized $N$ turns out to achieve the maximum fidelity. With the same $\gamma/g_a=10^{-2.5}$, the fidelity for generating $(|00\rangle\pm|33\rangle)/\sqrt{2}$ attains $\mathcal{F}\approx 94\%$ when $N=21$ and that for $(|00\rangle\pm|55\rangle)/\sqrt{2}$ attains $\mathcal{F}\approx 86\%$ when $N=29$. Over the peak value, the fidelity decreases with $N$. The success probabilities for $n=3$ and $5$, are shown in Figs.~\ref{2modegenesuperdecay}(c) and \ref{2modegenesuperdecay}(d), respectively. A larger $n$ gives rise to a lower final success probability.

\section{Conclusion}\label{conclusion}

In summary, we have constructed a protocol based on a sequence of projective measurements on the ancillary atomic system to generate Fock states and superposed Fock states from a coherent state of a resonator. The measurements are separated by dozens of rounds of joint free evolutions of the resonator and the ancillary system, which are coupled with a JC interaction. Through analysis of the population-reduction factors induced by the effective time-evolution operator for the target resonator, we found optimal strategies with varying duration of the evolution-and-measurement cycles. In the absence of decoherence, our protocol can therefore use fewer than $30$ measurements to create a Fock state $|n\rangle$ or a superposed Fock state $(|0\rangle+|n\rangle)/\sqrt{2}$ of $n\sim 10$ with a close-to-unit fidelity. Moreover, it can be generalized to prepare a Bell-like state $(|00\rangle+|nn\rangle)/\sqrt{2}$ in a double-resonator system assisted by a qutrit. Our protocol does not require any elaborate parametric control and external driving on the system. The success probability of our conditional protocol is determined by the overlap between the initial state and the target state of the resonator. In the presence of decoherence, the attainable fidelity for a given target state is limited by the decay to the lower states and an optimized number of measurements emerges for generating a Bell-like state. Our work thus proves quantum measurement to be a powerful tool to manipulate composite systems and presents a path to the generation of high-level Fock states and entangled states.

\section*{Acknowledgments}

We acknowledge financial support from the National Natural Science Foundation of China (Grant No. 11974311).

\bibliographystyle{apsrevlong}
\bibliography{ref}

\end{document}